# Title: Giant Tunneling Magnetoresistance in Spin-Filter van der Waals Heterostructures

**Authors:** Tiancheng Song,[1†] Xinghan Cai,[1†] Matisse Wei-Yuan Tu,[2] Xiaoou Zhang,[3] Bevin Huang,[1] Nathan P. Wilson,[1] Kyle L. Seyler,[1] Lin Zhu,[4] Takashi Taniguchi,[5] Kenji Watanabe,[5] Michael A. McGuire,[6] David H. Cobden,[1] Di Xiao,[3]* Wang Yao,[2]* Xiaodong Xu[1,4]*

**Affiliations:**
[1]Department of Physics, University of Washington, Seattle, Washington 98195, USA.
[2]Department of Physics and Center of Theoretical and Computational Physics, University of Hong Kong, Hong Kong, China.
[3]Department of Physics, Carnegie Mellon University, Pittsburgh, Pennsylvania 15213, USA.
[4]Department of Materials Science and Engineering, University of Washington, Seattle, Washington 98195, USA.
[5]National Institute for Materials Science, Tsukuba, Ibaraki 305-0044, Japan.
[6]Materials Science and Technology Division, Oak Ridge National Laboratory, Oak Ridge, Tennessee 37831, USA.

*Correspondence to: xuxd@uw.edu, wangyao@hku.hk, dixiao@cmu.edu.

†These authors contributed equally to the work.

**Abstract:**
Magnetic multilayer devices that exploit magnetoresistance are the backbone of magnetic sensing and data storage technologies. Here we report novel multiple-spin-filter magnetic tunnel junctions (sf-MTJs) based on van der Waals (vdW) heterostructures in which atomically thin chromium triiodide ($CrI_3$) acts as a spin-filter tunnel barrier sandwiched between graphene contacts. We demonstrate tunneling magnetoresistance which is drastically enhanced with increasing $CrI_3$ layer thickness, reaching a record 19,000% for magnetic multilayer structures using four-layer sf-MTJs at low temperatures. These devices also show multiple resistance states as a function of magnetic field, suggesting the potential for multi-bit functionalities using an individual vdW sf-MTJ. Using magnetic circular dichroism measurements, we attribute these effects to the intrinsic layer-by-layer antiferromagnetic ordering of the atomically thin $CrI_3$. Our work reveals the possibility to push magnetic information storage to the atomically thin limit, and highlights $CrI_3$ as a superlative magnetic tunnel barrier for vdW heterostructure spintronic devices.

**Main Text:**

One of the great assets of two-dimensional (2D) materials is the ability they give to engineer artificial heterostructures without the need for lattice matching. They thus provide a unique platform for exploring emerging phenomena and device function at the designed atomic interfaces (*1*, *2*). However, magnetic memory and processing applications were out of reach in van der Waals (vdW) heterostructures before the recent appearance of suitable 2D magnetic materials (*3–10*). One of these is the magnetic insulator chromium triiodide ($CrI_3$), which in bilayer form has been found to possess a layered-antiferromagnetic ground state. Magneto-optical Kerr effect (MOKE) measurements suggest that the spins align ferromagnetically out-of-plane within each layer but antiferromagnetically between layers, resulting in vanishing net magnetization (Fig. 1A, left) (*3*).

This layered-antiferromagnetic ordering makes $CrI_3$ desirable for realizing atomically thin magnetic multilayer devices. When the magnetizations of the two layers in a bilayer are switched between anti-parallel (Fig. 1A, left) and parallel states (Fig. 1A, middle and right), giant tunneling magnetoresistance (TMR) is produced by the double spin-filtering effect (*11*, *12*). In general, spin filters, which create and control spin-polarized currents, are the fundamental element in magnetic multilayer devices, such as spin valves (*13–15*), magnetic tunnel junctions (MTJs) (*16–21*) and double spin-filter MTJs (sf-MTJs) (*11*, *12*). Compared with the existing magnetic multilayer devices that require different choices of (metallic or insulating) magnets and spacers, the layered-antiferromagnetic structure in bilayer $CrI_3$ avoids the need for fabricating separate spin filters with spacers. This guarantees sharp atomic interfaces between spin filters, crucial for achieving large sf-TMR.

An even more intriguing possibility arises if the intrinsic layered-antiferromagnetic structure of $CrI_3$ extends beyond the bilayer. In this case, every layer should act as another spin filter oppositely aligned in series, greatly enhancing the sf-TMR as the number of layers increases. The associated multiple magnetic states may also enable multi-bit encoding in an individual sf-MTJ device. Moreover, being insulators, atomically thin $CrI_3$ single crystals can be integrated into vdW heterostructures as tunnel barriers in place of the non-magnetic dielectric which is usually hexagonal boron nitride (hBN) (*22*, *23*) or transition metal dichalcogenides (*24*), adding magnetic switching functionality. The realization of such vdW heterostructure sf-MTJs could produce novel 2D magnetic interface phenomena (*25*) and enable spintronics components such as spin current sources and magnetoresistive random-access memory (MRAM) (*26*).

Here, we demonstrate the layered-antiferromagnetic nature of few-layer $CrI_3$ and the realization of vdW engineered sf-MTJs with extraordinarily large sf-TMR and potential multi-bit functionalities. Figure 1B shows the essential structure of the sf-MTJ, which consists of two few-layer graphene contacts separated by a thin $CrI_3$ tunnel barrier. We have made and investigated devices with bilayer, trilayer, and four-layer $CrI_3$. All measurements were carried out at a temperature of 2 K, unless otherwise specified.

We begin with the case of bilayer $CrI_3$. The inset of Fig. 1C is an optical micrograph of a device obtained by stacking exfoliated 2D materials using a dry transfer process in a glovebox (see supplementary materials S1.1). The $CrI_3$ sf-MTJ is sandwiched between two hexagonal boron nitride (hBN) flakes to avoid degradation. The tunneling junction area is less than ~1 µm² to avoid effects caused by lateral magnetic domain structures (*3*, *4*). Figure 1C shows the tunneling current ($I_t$) as a function of DC bias voltage (*V*) at selected magnetic fields ($\mu_0 H$) (see supplementary materials S1.2). Unlike in tunneling devices using non-magnetic hBN as the barrier (*22*, *23*), it has a strong magnetic field dependence. As shown in Fig. 1C, $I_t$ is much smaller at $\mu_0 H = 0$ T (purple trace) than it is in the presence of an out-of-plane field ($\mu_0 H_\perp$, red trace) or an in-plane field ($\mu_0 H_\parallel$, green trace). This magnetic-field-dependent tunneling current implies a spin-dependent tunneling probability related to the field-dependent magnetic structure of bilayer $CrI_3$.

To investigate the connection between the bilayer $CrI_3$ magnetic states and the magnetoresistance, we measured $I_t$ as a function of $\mu_0 H_\perp$ at a particular bias voltage (−290 mV). The green and orange curves in Fig. 2A correspond to decreasing and increasing magnetic fields, respectively. $I_t$ exhibits plateaus with two values, about −36 nA and −155 nA. The lower plateau is seen at low fields, while there is a sharp jump to the higher plateau when the magnetic field exceeds a critical value. We also employed reflective magnetic circular dichroism (RMCD) to probe the out-of-plane magnetization of the bilayer CrI$_3$ near the tunneling area. Figure 2B shows the RMCD signal as a function of $\mu_0 H_\perp$ under similar experimental conditions to the magnetoresistance measurements (see supplementary materials S1.3). The signal is small at low fields, corresponding to a layered-antiferromagnetic ground state (↑↓ or ↓↑), where the arrows indicate the out-of-plane magnetizations in the top and bottom layers respectively. As the magnitude of the field increases there is a step up to a larger signal corresponding to the fully spin-polarized states (↑↑ and ↓↓), consistent with earlier MOKE measurements on bilayer CrI$_3$ (*3*). Additional bilayer device measurements can be found in the supplementary materials S2.1.

A direct comparison of $I_t$ and RMCD measurements makes the following explanation of the giant sf-TMR compelling: in the ↑↓ or ↓↑ states at low field the current is small because spin-conserving tunneling of an electron through the two layers in sequence is blocked. The step in $I_t$ occurs when the magnetic field drives the bilayer into the ↑↑ and ↓↓ states and this block is removed. This is known as the double spin-filtering effect (*11*, *12*), and it can be modeled by treating the two monolayers as tunnel-coupled spin-dependent quantum wells (see supplementary materials S2.2).

We quantify the sf-TMR by $(R_{ap} - R_p)/R_p$, where $R_{ap}$ and $R_p$ are the resistances with anti-parallel and parallel spin alignment in bilayer CrI$_3$ respectively, measured at a given bias. Figure 2C shows the value of this quantity as a function of bias extracted from the $I_t$-$V$ curves in Fig. 1C. The highest sf-TMR achieved is 310% for magnetization fully aligned perpendicular to the plane and 530% for parallel alignment. The sf-TMR decreases as temperature increases and vanishes above the critical temperature at about 45 K (see supplementary materials S2.3) (*3*).

The fact that the sf-TMR for in-plane magnetization is larger than for out-of-plane implies anisotropic magnetoresistance, which is a common feature in ferromagnets (*27*) and is a sign of anisotropic spin-orbit coupling due to the layered structure of CrI$_3$. The sf-TMR is also peaked at a certain bias and asymmetric between positive and negative bias. These observations are similar to the reported double sf-MTJs based on EuS thin films, where the asymmetry is caused by the different thickness and coercive fields of the two EuS spin-filters (*12*). Likewise, our data imply that the device lacks up-down symmetry, possibly because the few-layer graphene contacts are not identical in thickness. This broken symmetry also manifests as tilting of the current plateaus (Fig. 2A) and the finite non-zero RMCD value (Fig. 2B) in the layered-antiferromagnetic states (see supplementary materials S2.4).

To further investigate magnetic anisotropy and the assignment of magnetic states in the bilayer, we measured $I_t$ as a function of in-plane magnetic field. As shown in Fig. 2D (black curve), $I_t$ is smallest at zero field, in the layered-antiferromagnetic state, and smoothly increases with the magnitude of the field. This behavior has a natural interpretation in terms of a spin-canting effect. Once the magnitude of $\mu_0 H_\parallel$ exceeds about 4 T, the spins are completely aligned with the in-plane field and $I_t$ saturates. Simulations of the canting effect to match the data (dashed purple curve) yield a magnetic anisotropy field of 3.8 T (see supplementary materials S2.5), much larger than the out-of-plane critical magnetic field of $\pm 0.6$ T seen in Fig. 2A. These results therefore both demonstrate and quantify a large out-of-plane magnetic anisotropy in bilayer CrI$_3$.

We next consider the trilayer case. Figures 3A and 3B show $I_t$ and RMCD, respectively, for a trilayer CrI$_3$ sf-MTJ as a function of out-of-plane field. There are four plateaus in the RMCD signal, at $-14\%$, $-5\%$, $5\%$ and $15\%$, the ratio between which is close to $-3:-1:1:3$. By analogy with the analysis of the ↑↓ and ↓↑ layered-antiferromagnetic states in the bilayer, we identify the trilayer ground state as ↑↓↑ or ↓↑↓ at zero field. We conclude that the interlayer coupling in trilayer CrI$_3$ is also antiferromagnetic, and the net magnetization in the ground state, and thus the RMCD value, is 1/3 of the saturated magnetization when the applied field fully aligns the three layers (see supplementary materials 2.4). The jumps in $I_t$ and RMCD in Figs. 3A and 3B are caused by the magnetization of an individual layer flipping, similar to what is seen in metallic layered-antiferromagnets (*28–31*).

We deduce that the low current plateau at small fields in Fig. 3A occurs because the two layered-antiferromagnetic states (↑↓↑ and ↓↑↓) of the trilayer function as three oppositely polarized spin filters in series. Large enough fields drive the trilayer into fully spin-polarized states, which enhances tunneling and gives the high current plateaus. Figure 3C shows the sf-TMR as a function of bias derived from the $I_t$-$V$ curves shown in the inset. The peak values are about 2,000% and 3,200% for magnetization fully aligned perpendicular and parallel to the plane, respectively, revealing a drastically enhanced sf-TMR compared to bilayer devices.

Increasing the CrI$_3$ thickness beyond three layers unlocks more complicated magnetic configurations. Figures 4A and B show $I_t$ and RMCD, respectively, for a four-layer device. There are multiple plateaus in each, signifying several magnetic configurations with different effects on the tunneling resistance. The small RMCD signal at low fields, below ~0.8 T, corresponds to the fully antiferromagnetic ground state, either ↑↓↑↓ or ↓↑↓↑. The fact that the RMCD is not zero (Fig. 4B) can be attributed to the asymmetry of the layers caused by the fabrication process, as in the bilayer case above (see supplementary materials S2.4). As expected, these fully antiferromagnetic states are very effective at blocking the tunneling current since they act as four oppositely polarized spin filters in series, explaining the very low current plateau at small fields in Fig. 4A. Applying a large enough field fully aligns the magnetizations of all the layers (↑↑↑↑ or ↓↓↓↓), producing the highest plateaus in both $I_t$ and RMCD. Figure 4C shows the sf-TMR as a function of bias extracted from the $I_t$-$V$ curves in the inset. The peak values are now about 8,600% and 19,000% for perpendicular and parallel field, respectively, representing a further enhancement of the sf-TMR compared to bilayer and trilayer cases.

The RMCD of four-layer CrI$_3$ also shows intermediate plateaus at about half the values in the fully aligned states (Figs. 4B, 4F and fig. S10), corresponding to magnetic states with half the net magnetization of the fully aligned states. There are four possible magnetic states for the positive field plateau: $M_+\{↑↓↑↑,↑↑↓↑,↓↑↑↑,↑↑↑↓\}$, the four time-reversal copies ($M_-$) being the negative field counterparts (Fig. 4D). The resulting spin filter configuration should then correspond to one layer polarized opposite to the other three.

Remarkably, in the range of fields where these 1:3 configurations occur the tunneling current displays multiple plateaus. The green curve in Fig. 4A shows three distinct intermediate $I_t$ plateaus.

Two are in the positive field corresponding to the same intermediate +18% RMCD plateau, and one is in the negative field range. The orange curve, sweeping in the opposite direction, is the time-reversal copy of the green one. We rule out domain structure effects as the cause of these extra plateaus by examining field-dependent RMCD maps of all three measured four-layer CrI$_3$ sf-MTJs, none of which showed any domains (see supplementary materials S2.6). In addition, the tunnel junction area is also quite small compared with the typical domain size of a few microns in CrI$_3$ (*3*, *4*).

Instead, these current plateaus probably originate from distinct magnetic states. Whereas the four states in $M_+$ are indistinguishable in RMCD due to the same net magnetization, the tunneling current is likely to be sensitive to the position of the one layer with minority magnetization. First, the ↓↑↑↑ and ↑↑↑↓ have only one current-blocking interface while ↑↓↑↑ and ↑↑↓↑ have two (green line between adjacent layers with opposite magnetizations shown in Fig. 4D). Second, the current flow direction as well as the possibly asymmetric few-layer graphene contacts may introduce distinct sf-TMR either between the ↓↑↑↑ and ↑↑↑↓ states or between the ↑↓↑↑ and ↑↑↓↑ states (see supplementary materials S2.2). However, to identify the specific magnetic states corresponding to the current plateaus will require a means to distinguish the individual layer magnetizations (see supplementary materials S2.7 for the complete 16 magnetic states).

The four-layer CrI$_3$ sf-MTJ with multiple resistance states points to the potential for using layered antiferromagnets for multi-bit functionality in an individual sf-MTJ. Figures 4E, 4F and S10 show $I_t$ and RMCD for two other four-layer CrI$_3$ sf-MTJs. They exhibit one or two intermediate plateaus, rather than the three observed in Fig. 4A. The sample dependence suggests that these intermediate states are sensitive to the environment of the CrI$_3$, such as the details of the contacts, implying potential tunability, for example by electrostatically doping the graphene contacts. One exciting future direction could be to seek electrically controlled switching between several different magnetoresistance states. Already the sf-TMR of up to 19,000% observed in four-layer devices is an order of magnitude larger than that of MgO-based conventional MTJs (*19–21*), and several orders of magnitude larger than achieved with existing sf-MTJs under similar experimental conditions (*12*). Although the demonstrated vdW sf-MTJs only work at low temperatures, these results highlight the potential of 2D magnets and their heterostructures for engineering novel spintronic devices with unrivaled performance (*32*, *33*).

**References:**

1. A. K. Geim, I. V. Grigorieva, Van der Waals heterostructures. *Nature* **499**, 419–425 (2013).

2. K. S. Novoselov, A. Mishchenko, A. Carvalho, A. H. Castro Neto, 2D materials and van der Waals heterostructures. *Science* **353**, aac9439 (2016).

3. B. Huang *et al.*, Layer-dependent ferromagnetism in a van der Waals crystal down to the monolayer limit. *Nature* **546**, 270–273 (2017).

4. D. Zhong *et al.*, Van der Waals engineering of ferromagnetic semiconductor heterostructures for spin and valleytronics. *Sci. Adv.* **3**, e1603113 (2017).

5. C. Gong *et al.*, Discovery of intrinsic ferromagnetism in two-dimensional van der Waals crystals. *Nature* **546**, 265–269 (2017).

6. M. A. McGuire, H. Dixit, V. R. Cooper, B. C. Sales, Coupling of crystal structure and magnetism in the layered, ferromagnetic insulator $CrI_3$. *Chem. Mater.* **27**, 612–620 (2015).

7. M.-W. Lin *et al.*, Ultrathin nanosheets of $CrSiTe_3$: a semiconducting two-dimensional ferromagnetic material. *J. Mater. Chem. C*. **4**, 315–322 (2016).

8. Y. Tian, M. J. Gray, H. Ji, R. J. Cava, K. S. Burch, Magneto-elastic coupling in a potential ferromagnetic 2D atomic crystal. *2D Mater.* **3**, 025035 (2016).

9. X. Wang *et al.*, Raman spectroscopy of atomically thin two-dimensional magnetic iron phosphorus trisulfide ($FePS_3$) crystals. *2D Mater.* **3**, 031009 (2016).

10. J.-U. Lee *et al.*, Ising-type magnetic ordering in atomically thin $FePS_3$. *Nano Lett.* **16**, 7433–7438 (2016).

11. D. C. Worledge, T. H. Geballe, Magnetoresistive double spin filter tunnel junction. *J. Appl. Phys.* **88**, 5277–5279 (2000).

12. G.-X. Miao, M. Müller, J. S. Moodera, Magnetoresistance in double spin filter tunnel junctions with nonmagnetic electrodes and its unconventional bias dependence. *Phys. Rev. Lett.* **102**, 076601 (2009).

13. M. N. Baibich *et al.*, Giant magnetoresistance of (001)Fe/(001)Cr magnetic superlattices. *Phys. Rev. Lett.* **61**, 2472–2475 (1988).

14. G. Binasch, P. Grünberg, F. Saurenbach, W. Zinn, Enhanced magnetoresistance in layered magnetic structures with antiferromagnetic interlayer exchange. *Phys. Rev. B*. **39**, 4828–4830 (1989).

15. B. Dieny *et al.*, Giant magnetoresistive in soft ferromagnetic multilayers. *Phys. Rev. B*. **43**, 1297–1300 (1991).

16. M. Julliere, Tunneling between ferromagnetic films. *Phys. Lett.* **54A**, 225–226 (1975).

17. J. S. Moodera, L. R. Kinder, T. M. Wong, R. Meservey, Large magnetoresistance at room temperature in ferromagnetic thin film tunnel junctions. *Phys. Rev. Lett.* **74**, 3273–3276 (1995).

18. T. Miyazaki, N. Tezuka, Giant magnetic tunneling effect in $Fe/Al_2O_3/Fe$ junction. *J. Magn. Magn. Mater.* **139**, L231-L234 (1995).

19. S. Yuasa, T. Nagahama, A. Fukushima, Y. Suzuki, K. Ando, Giant room-temperature magnetoresistance in single-crystal Fe/MgO/Fe magnetic tunnel junctions. *Nat. Mater.* **3**, 868–871 (2004).

20. S. S. P. Parkin *et al.*, Giant tunnelling magnetoresistance at room temperature with MgO (100) tunnel barriers. *Nat. Mater.* **3**, 862–867 (2004).

21. S. Ikeda *et al.*, Tunnel magnetoresistance of 604% at 300 K by suppression of Ta diffusion in CoFeB/MgO/CoFeB pseudo-spin-valves annealed at high temperature. *Appl. Phys. Lett.* **93**, 082508 (2008).

22. L. Britnell *et al.*, Field-effect tunneling transistor based on vertical graphene heterostructures. *Science* **335**,

947-950 (2012).

23. G.-H. Lee *et al.*, Electron tunneling through atomically flat and ultrathin hexagonal boron nitride. *Appl. Phys. Lett.* **99**, 243114 (2011).

24. T. Georgiou *et al.*, Vertical field-effect transistor based on graphene–$WS_2$ heterostructures for flexible and transparent electronics. *Nat. Nanotechnol.* **8**, 100–103 (2012).

25. A. Soumyanarayanan, N. Reyren, A. Fert, C. Panagopoulos, Emergent phenomena induced by spin–orbit coupling at surfaces and interfaces. *Nature* **539**, 509–517 (2016).

26. S. A. Wolf *et al.*, Spintronics: a spin-based electronics vision for the future. *Science* **294**, 1488–1495 (2001).

27. T. R. Mcguire, R. I. Potter, Anisotropic magnetoresistance in ferromagnetic 3d alloys. *IEEE Trans. Magn.* **11**, 1018–1038 (1975).

28. O. Hellwig, T. L. Kirk, J. B. Kortright, A. Berger, E. E. Fullerton, A new phase diagram for layered antiferromagnetic films. *Nat. Mater.* **2**, 112–116 (2003).

29. O. Hellwig, A. Berger, E. E. Fullerton, Domain walls in antiferromagnetically coupled multilayer films. *Phys. Rev. Lett.* **91**, 197203 (2003).

30. B. Chen *et al.*, All-oxide–based synthetic antiferromagnets exhibiting layer-resolved magnetization reversal. *Science* **357**, 191–194 (2017).

31. M. Charilaou, C. Bordel, F. Hellman, Magnetization switching and inverted hysteresis in perpendicular antiferromagnetic superlattices. *Appl. Phys. Lett.* **104**, 212405 (2014).

32. D. C. Ralph, M. D. Stiles, Spin transfer torques. *J. Magn. Magn. Mater.* **320**, 1190–1216 (2008).

33. D. MacNeill *et al.*, Control of spin–orbit torques through crystal symmetry in $WTe_2$/ferromagnet bilayers. *Nat. Phys.* **13**, 300–305 (2016).

**Acknowledgements:** We thank Yongtao Cui and Guoxing Miao for insightful discussion. Work at the University of Washington was mainly supported by the Department of Energy, Basic Energy Sciences, Materials Sciences and Engineering Division (DE-SC0018171). Device fabrication and part of transport measurements are supported by NSF-DMR-1708419 and University of Washington Innovation Award. D.C.'s contribution is supported by DE-SC0002197. Work at CMU is supported by DOE BES DE-SC0012509. Work at HKU is supported by the Croucher Foundation (Croucher Innovation Award), Research Grant Council of HKSAR, and the HKU ORA. Work at ORNL (M.A.M.) was supported by the US Department of Energy, Office of Science, Basic Energy Sciences, Materials Sciences and Engineering Division. K.W. and T.T. acknowledge support from the Elemental Strategy Initiative conducted by the MEXT, Japan and JSPS KAKENHI Grant Numbers JP15K21722. D.X. acknowledges the support of a Cottrell Scholar Award. X.X. acknowledges the support from the State of Washington funded Clean Energy Institute and from the Boeing Distinguished Professorship in Physics.

**Figures:**

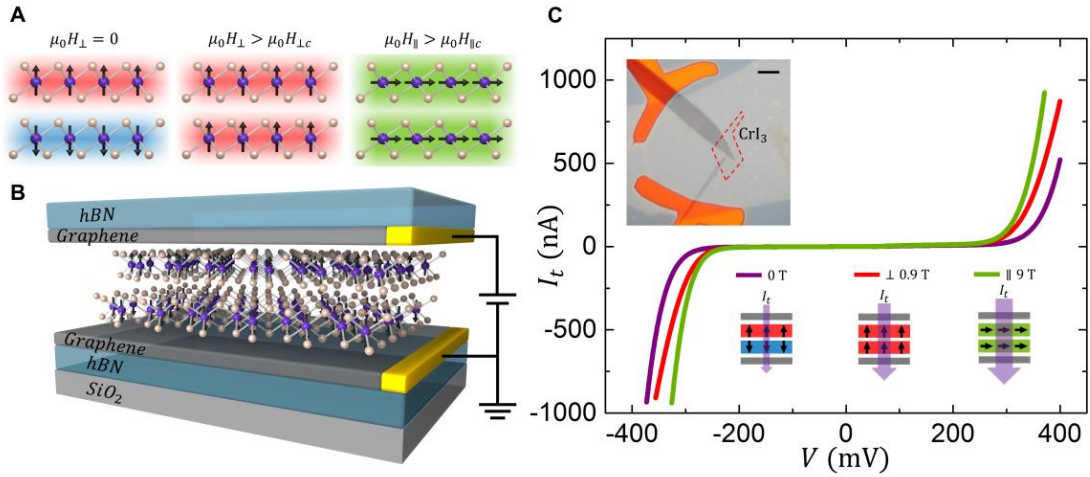

**Fig. 1. Spin-filter effects in layered-antiferromagnetic CrI$_3$.** (**A**) Schematic of magnetic states in bilayer CrI$_3$. Left: layered-antiferromagnetic state which blocks the tunneling current at zero magnetic field; middle and right: fully spin-polarized states with out-of-plane and in-plane magnetizations, which do not block it. (**B**) Schematic of 2D spin-filter magnetic tunnel-junction (sf-MTJ), with bilayer CrI$_3$ functioning as the spin-filter sandwiched between few-layer graphene contacts. (**C**) Tunneling current of a bilayer CrI$_3$ sf-MTJ at selected magnetic fields. Top inset: optical microscope image of the device (scale bar 5 μm). The red dashed line shows the position of the bilayer CrI$_3$. Bottom: schematic of the magnetic configuration for each $I_t$-$V$ curve.

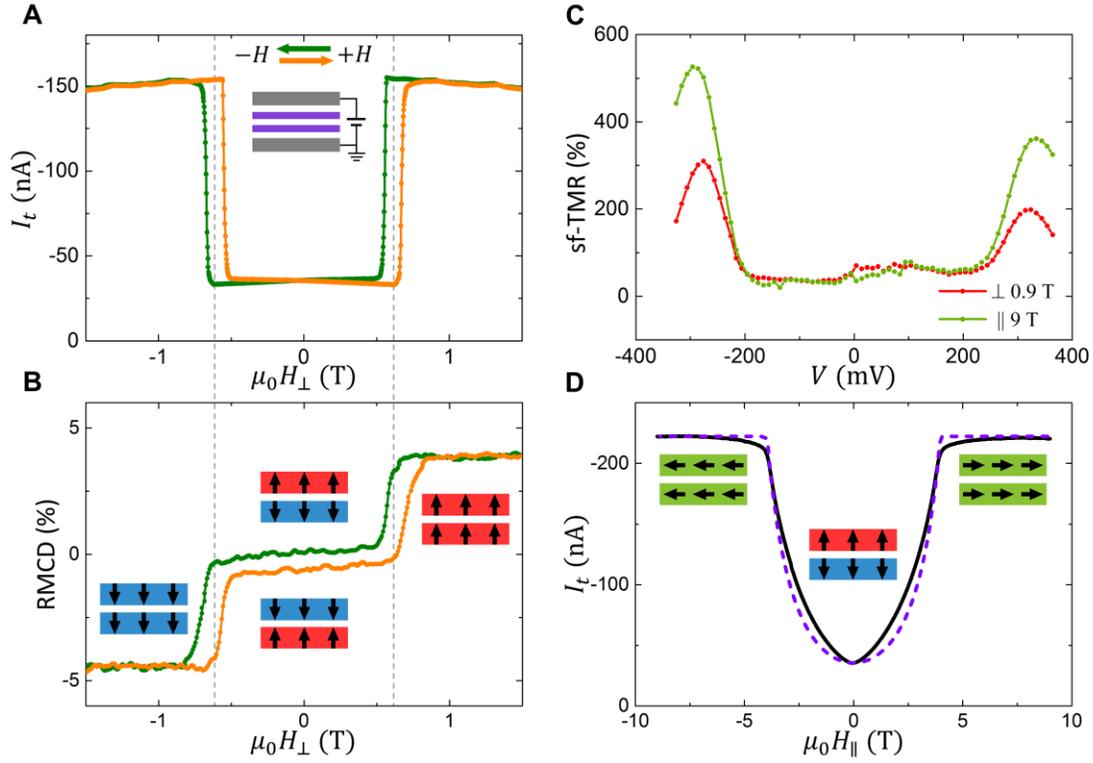

**Fig. 2. Double spin-filter MTJ from bilayer CrI$_3$.** (**A**) Tunneling current as a function of out-of-plane magnetic field ($\mu_0 H_\perp$) at a selected bias voltage ($-290$ mV). Green (orange) curve corresponds to decreasing (increasing) magnetic field. (**B**) Reflective magnetic circular dichroism (RMCD) of the same device at zero bias. Insets show the corresponding magnetic states. (**C**) Extracted sf-TMR ratio as a function of bias based on the $I_t$-$V$ curves in Fig. 1C. (**D**) Tunneling current as a function of in-plane magnetic field ($\mu_0 H_\parallel$) (black) at a selected bias voltage ($-290$ mV) with simulations (dashed purple). Insets show the corresponding magnetic states.

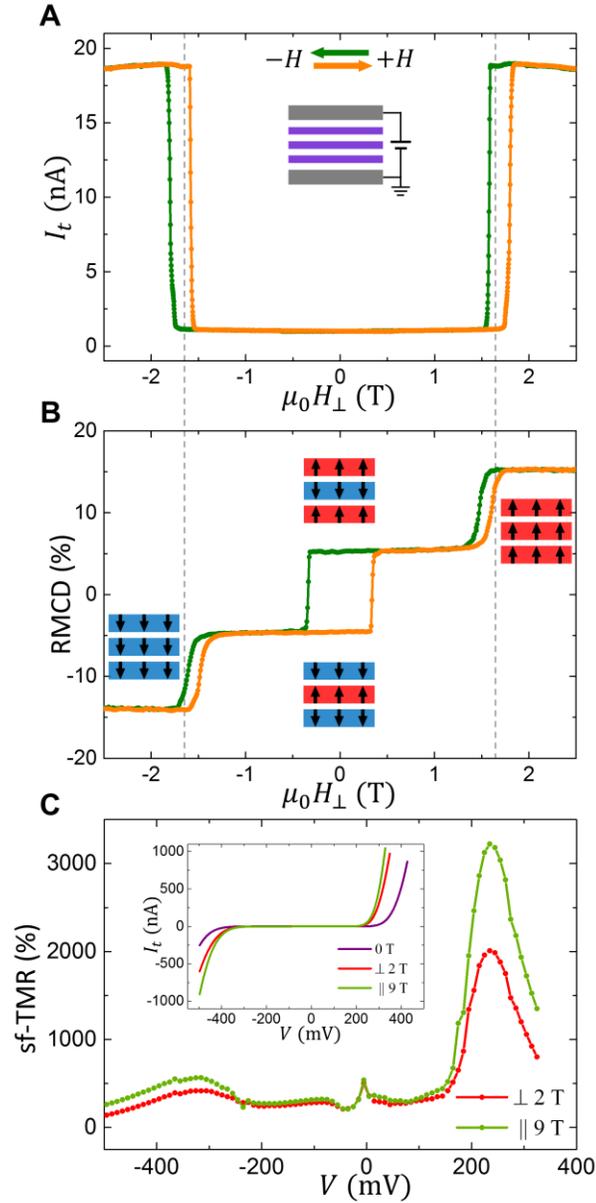

**Fig. 3. Giant sf-TMR of a trilayer CrI$_3$ sf-MTJ.** (**A**) Tunneling current as a function of out-of-plane magnetic field ($\mu_0 H_\perp$) at a selected bias voltage (235 mV). (**B**) RMCD of the same device at zero bias showing antiferromagnetic interlayer coupling. Insets show the corresponding magnetic states. (**C**) Calculated sf-TMR ratio from the $I_t$-$V$ data shown in the inset.

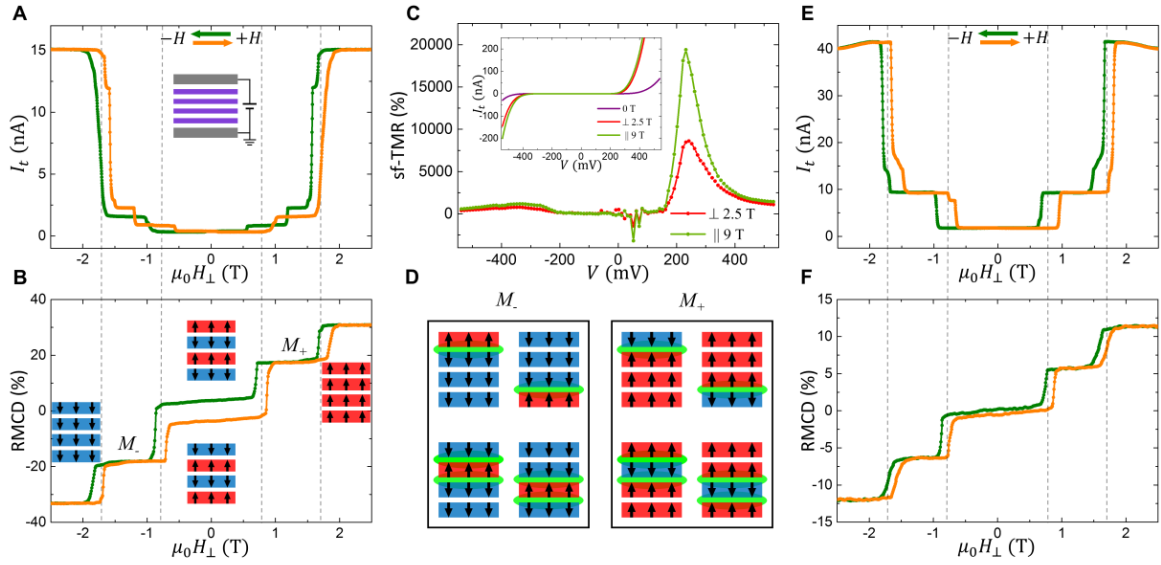

**Fig. 4. Four-layer CrI$_3$ sf-MTJs with extraordinarily large sf-TMR and multiple resistance states.** (**A**) Tunneling current as a function of out-of-plane magnetic field ($\mu_0 H_\perp$) at a selected bias voltage (300 mV), and (**B**) the corresponding RMCD of the same device at zero bias. Insets show the corresponding magnetic states. (**C**) Calculated sf-TMR ratio as a function of bias based on the $I_t$-$V$ curves in the inset. (**D**) Schematic of possible magnetic states corresponding to the intermediate plateaus in (A) and (B). Green lines show the current-blocking interfaces. (**E**) and (**F**) Tunneling current and RMCD from another four-layer CrI$_3$ sf-MTJ.

# Supplementary Materials for

## Giant Tunneling Magnetoresistance in Spin-Filter van der Waals Heterostructures

**Authors:** Tiancheng Song,[1†] Xinghan Cai,[1†] Matisse Wei-Yuan Tu,[2] Xiaoou Zhang,[3] Bevin Huang,[1] Nathan P. Wilson,[1] Kyle L. Seyler,[1] Lin Zhu,[4] Takashi Taniguchi,[5] Kenji Watanabe,[5] Michael A. McGuire,[6] David H. Cobden,[1] Di Xiao,[3]* Wang Yao,[2]* Xiaodong Xu[1,4]*

*Correspondence to: xuxd@uw.edu, wangyao@hku.hk, dixiao@cmu.edu.

†These authors contributed equally to the work.

**This PDF file includes:**

**S1. Materials and Methods**

**S1.1 Device fabrication**

**S1.2 Electrical measurement**

**S1.3 Reflective magnetic circular dichroism measurement**

**S2. Supplementary Text**

**S2.1 Additional bilayer $CrI_3$ sf-MTJs**

**S2.2 Tunneling through coupled magnetic quantum wells**

**S2.3 Temperature dependence of sf-TMR in bilayer $CrI_3$ sf-MTJs**

**S2.4 Monte Carlo simulations of layered-antiferromagnetic $CrI_3$**

**S2.5 Extraction of out of plane magnetic anisotropy**

**S2.6 RMCD maps of four-layer $CrI_3$ sf-MTJs**

**S2.7 Four-layer $CrI_3$ magnetic states**

**S2.8 Additional four-layer $CrI_3$ sf-MTJs**

**Figs. S1 to S10**

**Table S1**

## S1. Materials and Methods

### S1.1 Device fabrication

The hBN and few-layer graphene flakes were mechanically exfoliated onto $SiO_2$/Si wafers under ambient conditions and examined by optical and atomic force microscopy. Atomically smooth flakes were identified with thickness of 5-30 nm (hBN) and 2-8 nm (graphene). $CrI_3$ crystals were exfoliated inside an inert gas glovebox with oxygen and water levels below 1 ppm. The $CrI_3$ layer thickness was confirmed by comparing optical contrast of the microscope image to established optical contrast models of $CrI_3$ (*1*, *2*). V/Au (4/40 nm) metal electrodes were deposited onto the bottom hBN flake by electron beam evaporation following a standard electron-beam lithography technique with a bilayer resist (A4 495 and A4 950 poly (methyl methacrylate) (PMMA))). The top hBN and graphene were picked up using a polymer-based dry transfer technique (*3*). The $CrI_3$ flakes were then picked up and stacked between the top hBN/graphene and the bottom graphene/hBN in the glove box. In the resulting heterostructure, the $CrI_3$ flake is fully encapsulated, and the top/bottom graphene flakes are contacted by the pre-patterned metal electrodes. Finally, the polymer was dissolved in chloroform outside of the glovebox for less than one minute to minimize the exposure to ambient conditions.

### S1.2 Electrical measurement

The magneto-transport measurements were performed in a PPMS DynaCool cryostat (Quantum Design, Inc) with a base temperature 1.8 K and magnetic field up to 9 T. The $CrI_3$ sf-MTJ devices were mounted in a Horizontal Rotator probe, which allows device rotations around an axis perpendicular to the magnetic field of the longitudinal PPMS magnet. The schematic of $CrI_3$ sf-MTJs is shown in Fig. 1B. The DC bias voltage (*V*) is applied to the top graphene contact while the bottom graphene contact is grounded. The resulting tunneling current ($I_t$) is amplified and acquired by a current preamplifier (DL Instruments; Model 1211) and a DAQ device (National Instruments).

### S1.3 Reflective magnetic circular dichroism measurement

When a sample is magnetized with an out-of-plane magnetization *M*, the material may exhibit both magnetic circular birefringence (MCB) and magnetic circular dichroism (MCD). MCB induces a phase difference between right-circularly polarized (RCP) and left-circularly polarized (LCP) light while MCD imparts an amplitude difference between RCP and LCP light, with both effects resulting from *M*. When linearly polarized light, an equal superposition of RCP and LCP light, is reflected off the surface of the magnetized material, the phase difference between RCP and LCP light causes a rotation in the linear polarization, $\theta_K$, from the magneto-optical Kerr effect (MOKE), while the amplitude difference between RCP and LCP light induces elliptical polarization from reflective magnetic circular dichroism (RMCD).

In practice, RMCD measurements were performed in a closed-cycle helium cryostat (attoDRY 2100, attocube systems AG) with a base temperature of 1.55 K. A power-stabilized 632.8 nm HeNe laser was used to probe the sample at normal incidence with a fixed power of 30 μW. Magnetic fields of up to 9 T perpendicular to the sample plane were applied. The experimental setup follows closely to that of the MOKE measurements performed previously on $CrI_3$ – switching between MOKE and RMCD measurements only requires changing the lock-in frequency from the second harmonic of the photoelastic modulator (PEM) to the fundamental PEM frequency and removing the analyzing polarizer (*2*, *4*).

## S2. Supplementary Text

### S2.1 Additional bilayer $CrI_3$ sf-MTJs

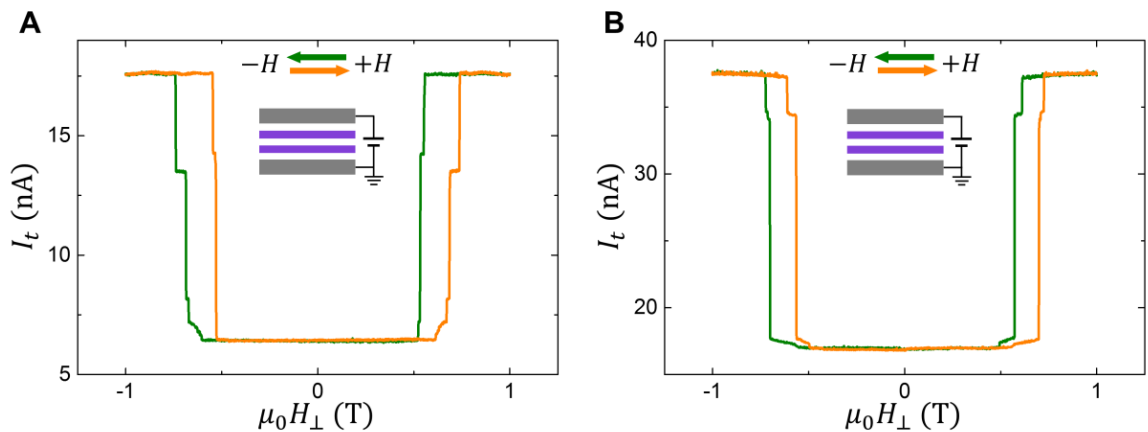

**Figure S1. Two additional bilayer $CrI_3$ sf-MTJs.** Tunneling current as a function of out-of-plane magnetic field ($\mu_0 H_\perp$) from two bilayer $CrI_3$ sf-MTJs (**A**, **B**) at a selected bias voltage (400 mV). Green (orange) curve corresponds to decreasing (increasing) magnetic field.

### S2.2 Tunneling through coupled magnetic quantum wells

With the strong vertical confinement of carriers in each monolayer and their weak vdW coupling in the heterostructures, we model the transport through few-layer $CrI_3$ sf-MTJ devices as coupled quantum wells as illustrated in fig. S2. Each monolayer with ferromagnetic ordering corresponds to a quantum well with spin-split subbands determined by the magnetization orientation. These magnetic quantum wells are separated from each other and from the graphene electrodes by barriers that correspond to the vdW gaps. Interlayer hopping across these barriers conserves spin, and also conserves momentum for hopping between the lattice-matched $CrI_3$ layers. Thus, the staggered quantum well states alignment in the layered-antiferromagnetic configuration leads to larger tunneling resistance compared to the fully spin-polarized case. The distinct efficiencies of transport through these two different quantum well state arrangements are readily anticipated at this level of modeling to give rise to significant sf-TMR.

The current is calculated using the Landauer-Buetikker formula,

$$I = \int \left(f_T(E) - f_B(E)\right) T(E) \, dE,$$

in which $f_l(E)$ is the Fermi distribution function of contact $l$ ($l$ =T for top and $l$ =B for bottom contact) and $T(E) = \sum_\sigma T_\sigma(E)$ is the total transmission with $T_\sigma(E)$ the transmission of carriers with spin $\sigma$ at incident energy $E$. At zero-bias, carriers with spin $\sigma$ experience a potential energy $V_0^\sigma(x)$ as a function of the out-of-plane coordinate $x$, describing the coupled quantum well profile dependent on the layered magnetic states (c.f. fig. S2). In general, the applied bias also produces an electric field across the multilayer tunnel barrier, imposing a linear gradient to the zero-bias potential profile $V_0^\sigma(x)$, resulting in the finite-bias potential profile: $V_\sigma(x) = V_0^\sigma(x) + \alpha V x$. The transmission $T_\sigma(E)$ is then calculated according to $V_\sigma(x)$ for each bias.

The calculation proceeds by discretizing the potential profile into a tight-binding Hamiltonian $H = \sum_\sigma H_\sigma$ with $H_\sigma = \sum_{i=-\infty}^{+\infty} V_\sigma(x_i) c_{i\sigma}^+ c_{i\sigma} + [-J \sum_{i=-\infty}^{+\infty} c_{i\sigma}^+ c_{i+1\sigma} + h.c.]$ describing the Hamiltonian for spin $\sigma$ and $x_i$ denoting the coordinate of the discretized site on lattice $i$. The potential profile is $V_\sigma(x_i) = \varepsilon_0$ for $i = -\infty, ..., 0$ and $i = D + 1, ..., \infty$. It is planned according to $V_\sigma(x_i)$ in the device region $i = 1, ..., D$ as shown in fig. S2 with the bias-produced modulation described above. The inter-site hopping in the tight-binding calculation depends on the discretization of the coordinate. Setting $J = 2$ eV corresponds to $x_{i+1} - x_i = 0.138$ nm, and we use $\varepsilon_0 = 0$ as a reference. We then solve the corresponding scattering-state problem for each incident energy $E$ for each spin species to obtain the transmission $T_\sigma(E)$.

The height of the vdW barriers is taken to be 4 eV, roughly the electron affinity of $CrI_3$. The spin-splitting $\Delta_0$ in each quantum well is taken to be 0.5 eV, as estimated from the DFT calculated band structure of the ferromagnetic monolayer $CrI_3$ (5). Typical $I_t$-$V$ relations produced from this model, with a van der Waals barrier width $B = 0.276$ nm and quantum well width $W = 0.552$ nm, are shown in fig. S3A-C. For this choice of parameters, the contact Fermi energy at zero bias is ~ 0.4 eV below the lowest quantum well bands. The turn-on behavior of the current is qualitatively well represented. All these calculations are done with a constant $\alpha = 0.05$, under the assumption that the electric field in the $CrI_3$ layers is linearly proportional to the bias. The sf-TMR ratio, as shown in fig. S3D, increases with the number of layers.

We also compared the currents of various layered magnetic states of four-layer devices at an exemplified bias $V = 0.3$ eV, in Table S1. These results are consistent with the observation that the states with larger net magnetizations have larger currents. Moreover, the states with the same magnetizations (e.g. the 1:3 magnetic states discussed in the main text) also correspond to different currents, although the magnitude of the differences depends on the parameters $W$ and $B$.

This calculation treats $CrI_3$ in different magnetic states as rigid scattering potentials, which can be oversimplified, especially in the large bias regime when the current becomes large. Various possible processes including the spin relaxation, relaxation of in-plane momentum (e.g. due to defects) in hopping between the $CrI_3$ layers, and magnetization dynamics caused by the current, cannot be properly accounted due to a lack of detailed information. The simplified description of the ferromagnetic monolayer as a spin-dependent quantum well of square shape, and the vdW gap by a square tunnel barrier, also results in some uncertainty in determining the width of the well and the thickness of the barrier, as both are already in the atomically thin limit. Moreover, in the $I_t$-$V$ curves calculations, we have assumed that the electric field across the $CrI_3$ grows linearly with the applied bias voltage. This assumption can also break down when the current becomes large, where a self-consistent determination of the steady-state carrier distribution and hence the electric field would be necessary. Nevertheless, the dramatic enhancement of the sf-TMR with the increase of layer number, and the distinguishability by the tunneling current of different magnetic states with the same net magnetization, can be captured here.

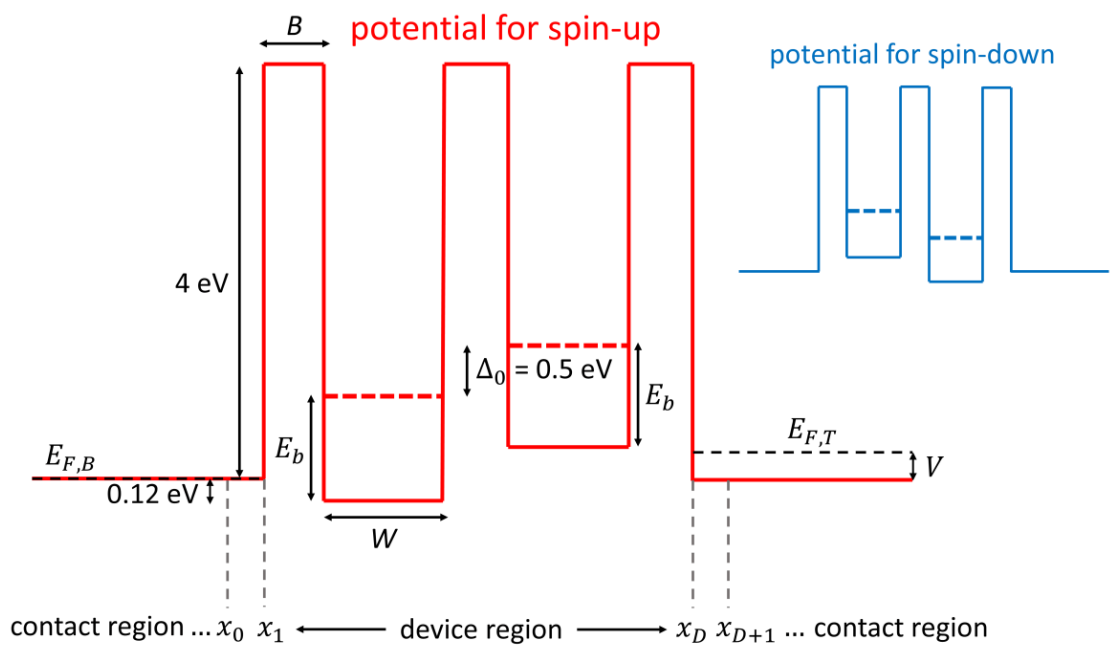

**Figure S2. Schematic plot of the potential profile for the model of coupled magnetic quantum wells**. The red solid line sketches the potential experienced by spin up carriers in the layered-antiferromagnetic state of bilayer system, while the one for spin down carriers is schematically shown in the inset by the blue solid line. We denote the barrier thickness by $B$, and the well width by $W$. The red (blue) dashed horizontal lines indicate the edge of the lowest spin-up (spin-down) subbands in the two quantum wells. The splitting between the spin-up and spin-down subbands is denoted by $\Delta_0$, taken to be 0.5 eV here. For the quantum well width $W$ = 0.414 nm and 0.552 nm used in the calculations of fig. S3 and Table S1, the quantum confinement energies $E_b \approx \frac{2(\hbar)^2}{mW^2}$ in the vertical direction are estimated to be 0.89 eV and 0.5 eV respectively, so both spin subbands in the quantum wells are above the contact Fermi levels, $E_{F,T}$ (for the top) and $E_{F,B}$ (for the bottom, grounded) in the applied bias range $V \leq 0.4$ V. The spatial coordinate is discretized in the tight-binding calculation (c.f. text). The device region and the contact regions are delineated as indicated at the bottom of this figure.

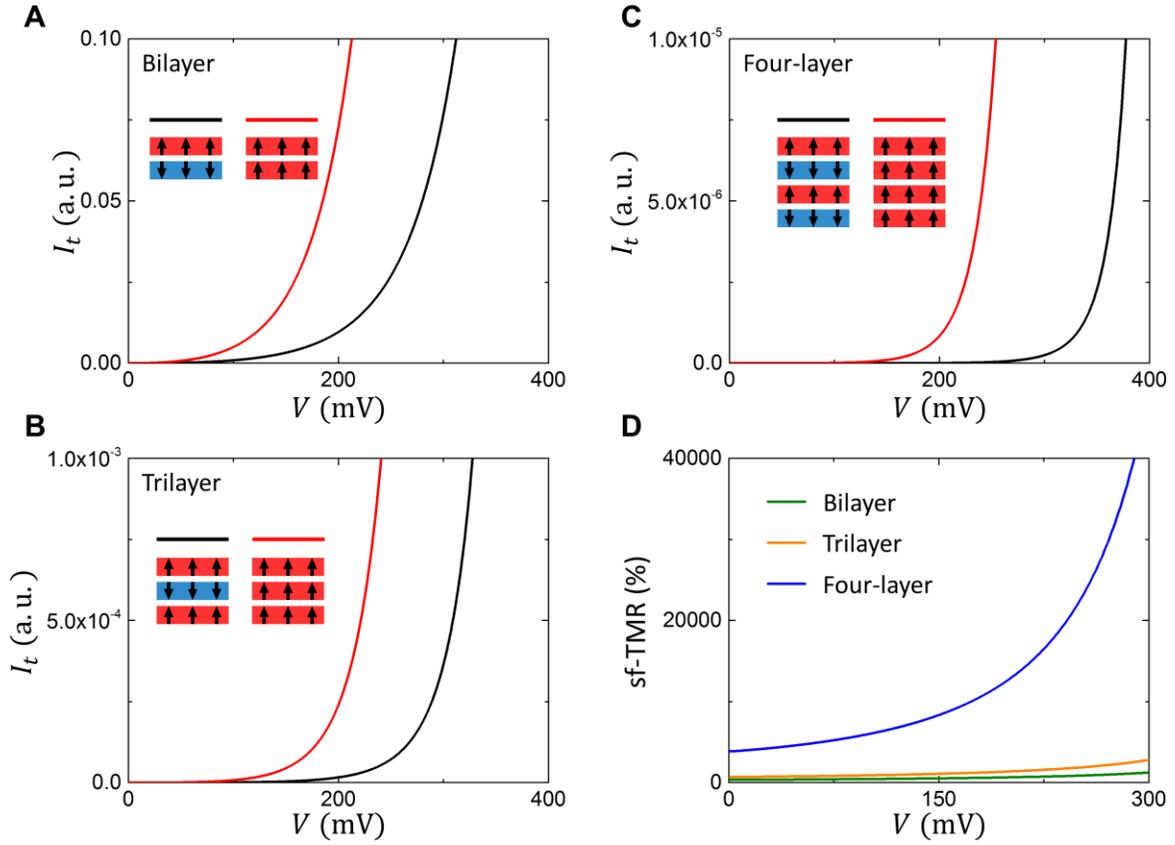

**Figure S3.** (**A**), (**B**) and (**C**) Calculated $I_t$-$V$ curves for different number of layers as indicated. (**D**) The sf-TMR ratios for bilayer (green), trilayer (orange) and four-layer (blue). Other parameters are well width $W = 0.552$ nm and barrier thickness $B = 0.276$ nm.

|  | ↑↑↑↑ | ↑↑↓↑ | ↑↓↑↑ | ↑↑↑↓ | ↓↑↑↑ | ↑↑↓↓ | ↑↓↓↑ |
|---|---|---|---|---|---|---|---|
| $B$=0.138nm $W$=0.414nm | 31.1231 | 3.86883 | 3.85506 | 4.18626 | 4.02135 | 1.06243 | 1.03059 |
| $B$=0.138nm $W$=0.552nm | 3923.15 | 28.8291 | 27.9247 | 53.151 | 46.2536 | 1.46031 | 1.23217 |
| $B$=0.276nm $W$=0.552nm | 484.588 | 15.2186 | 15.1638 | 15.626 | 15.3802 | 1.01942 | 1.00968 |

**Table S1.** The ratio of currents of various magnetic states (indicated on the first row) to that of the ↑↓↑↓ state. $W$ denotes the width of the wells and $B$ denotes the thickness of the van der Waals barriers (c.f. fig. S2).

## S2.3 Temperature dependence of sf-TMR in bilayer CrI₃ sf-MTJs

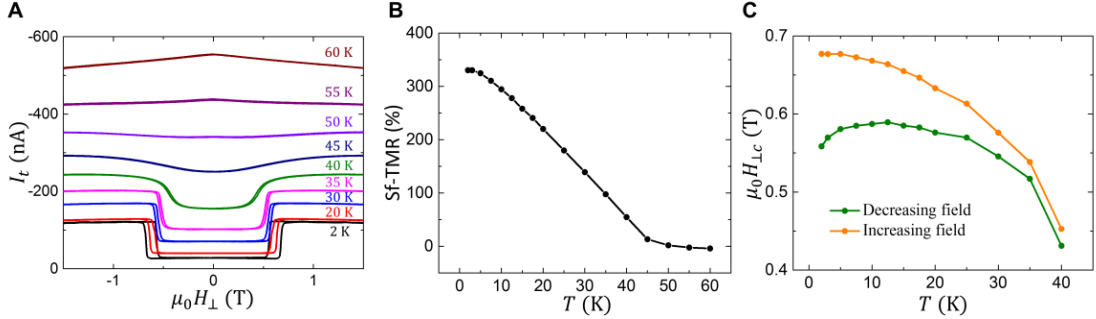

**Figure S4. Temperature dependence of sf-TMR in the bilayer CrI₃ sf-MTJ shown in Fig. 2A.** (**A**) Temperature dependence of tunneling current as a function of out-of-plane magnetic field ($\mu_0 H_\perp$) at a selected bias voltage ($-285$ mV). (**B**) The sf-TMR decreases as temperature increases and finally vanishes around the critical temperature of few-layer CrI₃. (**C**) Temperature dependence of the spin-flip transition critical magnetic field ($\mu_0 H_{\perp c}$), which shows a trend of shrinking and shifting of the hysteresis loop. The two critical fields ($\mu_0 H_{\perp c}$) are defined at decreasing and increasing positive magnetic fields.

## S2.4 Monte Carlo simulations of layered-antiferromagnetic CrI₃

The magnetic structure of CrI₃ in the few-layer limit is layered-antiferromagnetic with out-of-plane magnetic anisotropy. This can be described by the Ising model, where the spins in each layer have ferromagnetic intralayer coupling while the interlayer coupling is antiferromagnetic. The magnetizations of bilayer, trilayer and four-layer CrI₃ shown in Figs. 2B, 3B and 4B in the main text, respectively, are in good agreement with the Monte Carlo simulations performed for this Ising model (*6*).

We simulated the CrI₃ magnetic structure by choosing the Ising model based on CrI₃ crystal structure, using the Monte Carlo method with classical Ising spins ($S = \pm 1$). The Ising model consists of $N$ ferromagnet layers, where each layer is a $L \times L$ honeycomb lattice with one spin at each lattice site, corresponding to each Cr atom site. The intralayer nearest-neighbor interactions are ferromagnetic, corresponding to a positive exchange coupling ($J_{intra}$). The antiferromagnetic interlayer interaction is implemented by a negative exchange coupling ($J_{inter}$) between nearest-neighbor sites in adjacent layers. The Hamiltonian of the Ising model is defined as

$$\mathcal{H} = \sum_{n=1}^{N} (-\frac{1}{2} J_{intra} \sum_{<i,j>} S_{n,i} S_{n,j} - \frac{1}{2} J_{inter} \sum_{i} S_{n,i} S_{n\pm 1,i} - H \sum_{i} S_{n,i}),$$

where $S_{n,i}$ is the spin on the $i$ site of the $n$ layer, and $H$ is the external magnetic field.

The magnetization as a function of $H$ was simulated using a Metropolis Monte Carlo algorithm by flipping the individual spin on each site one-by-one. We simulated a $50 \times 50$ honeycomb lattice with periodic boundary conditions. We set $J_{intra} = -J_{inter} = 1$ and express the external magnetic field ($H$) in the unit of $J_{inter}$. We chose to simulate at a selected temperature ($T \approx 0.5\, T_c$) to allow the spins to fluctuate and avoid trapping in a local energy-minimum state. $2000 \times N \times L \times L$ Monte Carlo steps were performed to make sure the system reached a quasi-equilibrium state at each magnetic field, then the net magnetization was calculated.

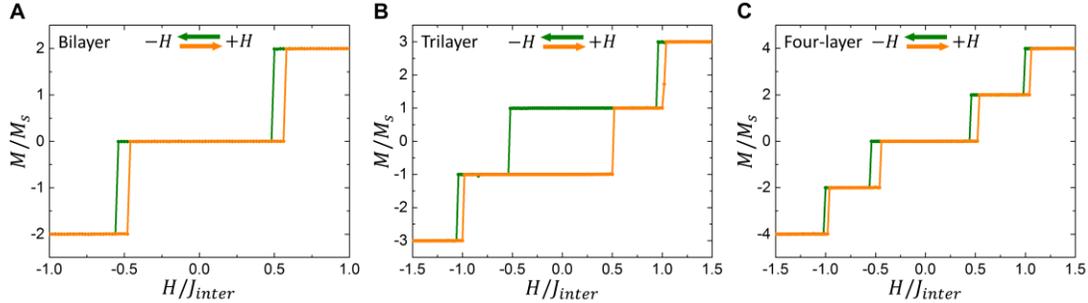

**Figure S5. Monte Carlo simulations of symmetric CrI$_3$.** All layers of bilayer (**A**), trilayer (**B**) and four-layer (**C**) CrI$_3$ are assigned the same spin value $S = \pm 1$. Green (orange) curve corresponds to decreasing (increasing) magnetic field. $M_s$ is the saturation magnetization of a monolayer CrI$_3$ with spin values $S = \pm 1$.

Figures S5A, B and C show the simulated magnetizations of bilayer, trilayer and four-layer CrI$_3$, respectively, which are in good agreement with the RMCD data shown in Figs. 2B, 3B and 4B, except for the finite non-zero magnetizations at zero field in bilayer and four-layer CrI$_3$. This is because in the simulations, it is assumed that the spins of each layer are identical, and thus have the same spin value $S = \pm 1$. However, in the experiment, the symmetry between the top and bottom layers of CrI$_3$ can be broken during the heterostructure fabrication, such as the asymmetric top and bottom hBN and few-layer graphene contacts. To simulate an asymmetric bilayer CrI$_3$, we slightly change the spin values in the top and bottom CrI$_3$ layers to break the symmetry. Figure S6A shows the simulated magnetization of an asymmetric bilayer CrI$_3$ with spin values $S = \pm 1.05$ and $S = \pm 0.95$ for the top and bottom layers respectively. Similar simulated magnetization of an asymmetric four-layer CrI$_3$ is shown in fig. S6B. The finite non-zero simulated magnetizations of the asymmetric bilayer and four-layer CrI$_3$ agree well with the RMCD data shown in Figs. 2B and 4B, supporting the assumption of symmetry breaking in CrI$_3$. The tilted current can be explained in the similar way, as the magnetic field can affect the symmetry breaking, thus modify the spin-filtering efficiency of the ↑↓ and ↓↑ magnetic ground states.

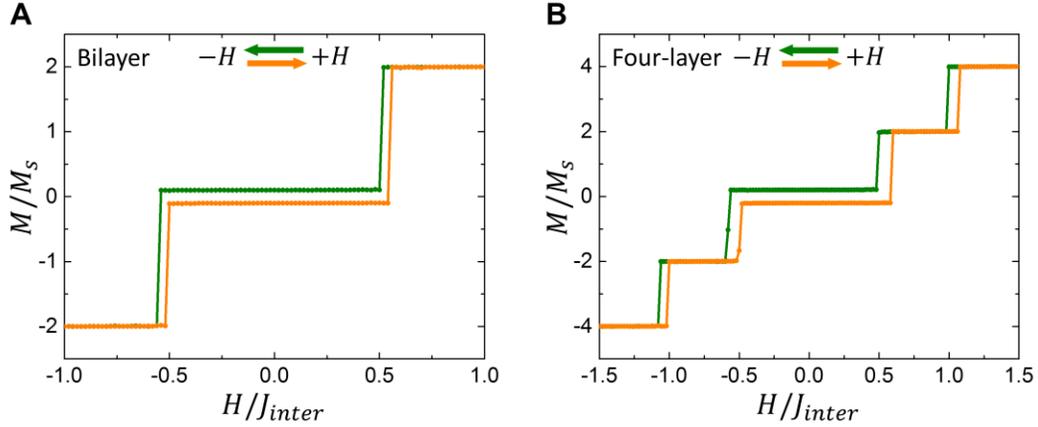

**Figure S6. Monte Carlo simulations of asymmetric CrI$_3$.** (**A**) Bilayer CrI$_3$ is assigned spin values $S = \pm 1.05$ and $S = \pm 0.95$ for the top and bottom layers respectively. (**B**) Four-layer CrI$_3$ is assigned with spin values $S = \pm 1.1$, $S = \pm 1$, $S = \pm 1$ and $S = \pm 0.9$ for the four layers from top to bottom respectively. In both the bilayer and the four-layer, the different spin amplitudes between each layer – equivalent to breaking out-of-plane inversion symmetry – results in the appearance of spontaneous magnetization.

### S2.5 Extraction of out of plane magnetic anisotropy

The magnetic anisotropy and the interlayer exchange interaction can be estimated by fitting the tunneling current with a numerical model which will be discussed in this section. There are two contributing effects to the tunneling current in the CrI$_3$ sf-MTJs: the anisotropic magnetoresistance of each layer and the spin-dependent tunneling at the interface between adjacent CrI$_3$ layers.

The relation between the magnetoresistance of each layer and the direction of magnetization can be achieved from symmetry analysis. Under all symmetry operations, the resistance tensor $\rho_{ij}(\boldsymbol{\alpha})$ should be invariant, where $\boldsymbol{\alpha}$ is the direction of the magnetization. Specifically, we expand the resistance tensor into a power series of $\boldsymbol{\alpha}$ (7)

$$\rho_{ij}(\boldsymbol{\alpha}) = c_{ij} + c_{ijk}\alpha_k + c_{ijkl}\alpha_k \alpha_l \ldots .$$

Under a symmetry operation $T_{ij}$, the following relations stand

$$c_{ij} = T_{ii'}T_{jj'}c_{i'j'},$$

$$c_{ijkl} = T_{ii'}T_{jj'}T_{kk'}T_{ll'}c_{i'j'k'l'}.$$

In the following discussion, we concentrate on the longitudinal resistance $\rho_{33}$ along the $z$ direction, and only keep the expansion terms up to the second order. According to the Onsager reciprocal relation $\rho_{ij}(\boldsymbol{\alpha}) = \rho_{ji}(-\boldsymbol{\alpha})$ (8), the coefficients for the odd-order terms are antisymmetric with respect to the $i$ and $j$ indices. Therefore, these antisymmetric terms have no contributions to the longitudinal resistance, and only the even-order terms survive.

Using the symmetry group of CrI3, i.e. $D_{3d}$, the only nonzero coefficients related to $\rho_{33}$ are $c_{33}$, $c_{3311} = c_{3322}$ and $c_{3333}$. Resultantly, the resistance along the *z* direction is given by

$$\rho_{33}(\theta_M) = c_{33} + \frac{1}{2}(c_{3333} + c_{3311}) + \frac{1}{2}(c_{3333} - c_{3311})\cos(2\theta_M) = \rho_0 + \rho_1 \cos(2\theta_M),$$

where $\theta_M$ is the angle between $\boldsymbol{\alpha}$ and the *z* axis, and $\rho_0$, $\rho_1$ are the angular dependent and independent resistances, respectively. According to our experimental results, the tunneling current reaches its maximum when the magnetic field is in-plane, which implies that $\rho_1$ is positive. This cannot be guaranteed from the symmetry analysis. Notice that the direction of magnetization $\theta_M$ is different from the direction of the magnetic field $\theta_B$. $\theta_M$ can be calculated by minimizing the energy of the bilayer system

$$H = J\boldsymbol{s}_1 \cdot \boldsymbol{s}_2 - \frac{K}{2}(\boldsymbol{s}_{1z}^2 + \boldsymbol{s}_{2z}^2) - \boldsymbol{B} \cdot (\boldsymbol{s}_1 + \boldsymbol{s}_2).$$

Here $J > 0$ is the interlayer antiferromagnetic coupling, $K$ is the anisotropy, and $\boldsymbol{s}_{1,2}$ are the spins of the two layers. In this simple model, we assume that the spins within the same layer are aligned, such that the intralayer exchange coupling term is just a constant.

Apart from the anisotropic resistance within each layer, the spin-dependent tunneling at the interface of two adjacent layers also affects the tunneling current. Suppose the spins of both layers are in the *x-z* plane, and the angles between the spin polarizations and the *z* axis are $\theta_{M1}$ and $\theta_{M2}$, respectively. The spin dependent tunneling coefficient can be decomposed into spin-parallel and antiparallel tunneling coefficients as

$$t(\theta_{M1}, \theta_{M2}) = |\langle\theta_{M1}|\theta_{M2}\rangle|^2 t_P + (1 - |\langle\theta_{M1}|\theta_{M2}\rangle|^2)t_{AP}$$
$$= \cos^2\frac{\theta_{M1} - \theta_{M2}}{2} t_P + \sin^2\frac{\theta_{M1} - \theta_{M2}}{2} t_{AP},$$

where $|\theta\rangle = (\cos\theta/2, \sin\theta/2)^T$ is the spin wave function, $t_P$ and $t_{AP}$ are the tunneling coefficients for the spin-parallel and antiparallel cases, respectively. Combining both the anisotropic magnetoresistance and the spin-dependent tunneling, the conductance of a bilayer structure is given by

$$\sigma_{\theta_{M1}, \theta_{M2}} = \frac{t(\theta_{M1}, \theta_{M2})}{2\rho_0 + \rho_1 \cos(2\theta_{M1}) + \rho_1 \cos(2\theta_{M2})}.$$

This expression can be easily generalized to trilayer and four-layer cases.

The values of all unknown parameters can be estimated by numerically fitting the tunneling current. Two different measurements are used for the fitting: rotating a 9 T magnetic field in the *x-z* plane (shown in fig. S7), and increasing the magnitude of an in-plane magnetic field from zero to 9 T (shown in Fig. 2D). In the first experiment, the spins of all layers are aligned, and we can get $\rho_0$, $\rho_1$, and $K$ according to the shape and magnitude of the angular-dependent tunneling current. In the second experiment, the system is in the layered-antiferromagnetic states when $B = 0$ T, while all spins are aligned when the magnetic field is large enough. Changing the relative direction between the spins makes it possible to estimate $J$, $t_P$ and $t_{AP}$. We note that the experimental result in fig. S7 shows more structures than the smooth numerical result. This is an indication that a more complex model is required to fully describe the magnetic property of the CrI3 sf-MTJs, which deserves further investigations in the future.

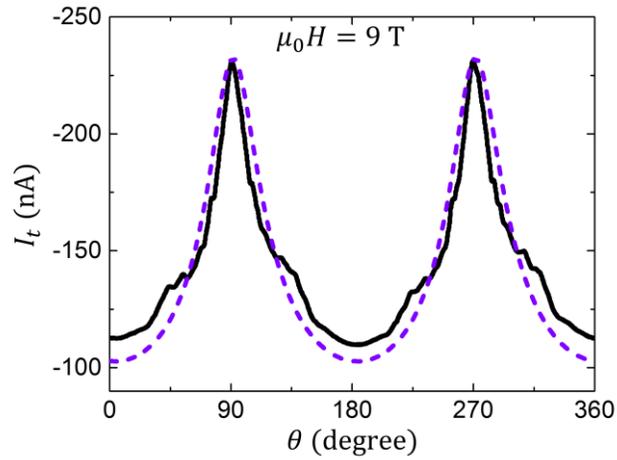

**Figure S7. Angular magnetoresistance of bilayer $CrI_3$.** Tunneling current as a function of the direction of 9 T magnetic field in the *x-z* plane ($\mu_0 H$) (black) at a selected bias voltage ($-290$ mV) with simulations (dashed purple). $\theta$ is the angle between the magnetic field and the out-of-plane direction. $\theta = 0°$ and $\theta = 90°$ correspond to out-of-plane and in-plane 9 T magnetic fields, respectively.

## S2.6 RMCD maps of four-layer CrI$_3$ sf-MTJs

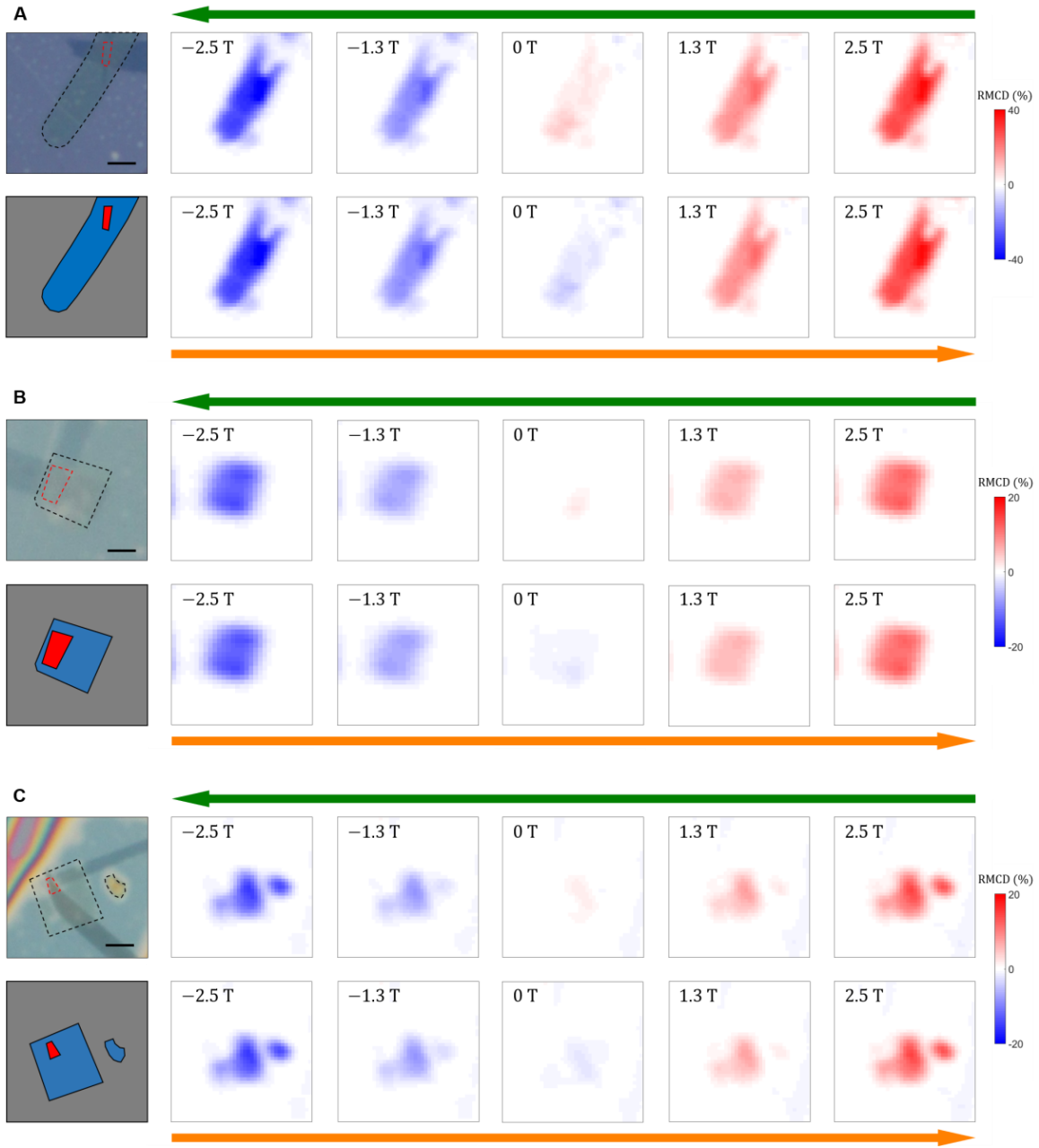

**Fig. S8. RMCD maps of three CrI$_3$ sf-MTJs at selected magnetic fields.** The green (orange) arrow corresponds to the decreasing (increasing) magnetic field. The first column shows the optical microscope image with the schematic of the device (scale bar 3 μm). The black dashed lines show the positions of the CrI$_3$. The red dashed lines show the positions of the overlap area (i.e. the junction area). (**A**), (**B**), and (**C**) correspond to sf-MTJs shown in Figs. 4A, 4E, and fig. S10, respectively.

Figure S8 shows RMCD maps at selected magnetic fields for three four-layers CrI$_3$ sf-MTJs, which rule out the possibility of domain effects. The RMCD maps at 1.3 T do not show domains of opposite or different magnetization. In addition, the RMCD maps at 1.3 T acquired from an upward and downward field sweep are identical, indicating that the field sweeping direction does not introduce domain effects at this intermediate magnetic field. However, as shown in Fig. 4A and fig. S10A, the tunneling current at 1.3 T has distinct plateaus when the magnetic field sweeps upwards versus downwards, implying the distinct intermediate magnetic states with the same net magnetization as the origin of different current plateaus. The same analysis can be made with RMCD maps at −1.3 T.

We note that there is slightly spatial inhomogeneity of RMCD signal across the devices as well as a difference in magnitude between devices. These effects likely arise for the following reasons. The CrI$_3$ flake is sandwiched between the top and bottom few-layer graphene contacts, which are further encapsulated by the top and bottom hBN flakes. The multilayer structure introduces thin-film interference from multiple reflections at the different interfaces of the heterostructure. Therefore, the difference in RMCD signal between devices is caused by different thickness of hBN flakes and few-layer graphene contacts. For each device, only a small part of the CrI$_3$ flake is sandwiched between two few-layer graphene contacts, and the overlap area (i.e. the junction area) is designed to be much smaller than the size of CrI$_3$ flake to avoid a short circuit and domain effects. The areas of CrI$_3$ with and without the few-layer graphene contacts differ in the magnitude of the RMCD signal, leading to the spatial non-uniformity in RMCD maps.

## S2.7 Four-layer CrI$_3$ magnetic states

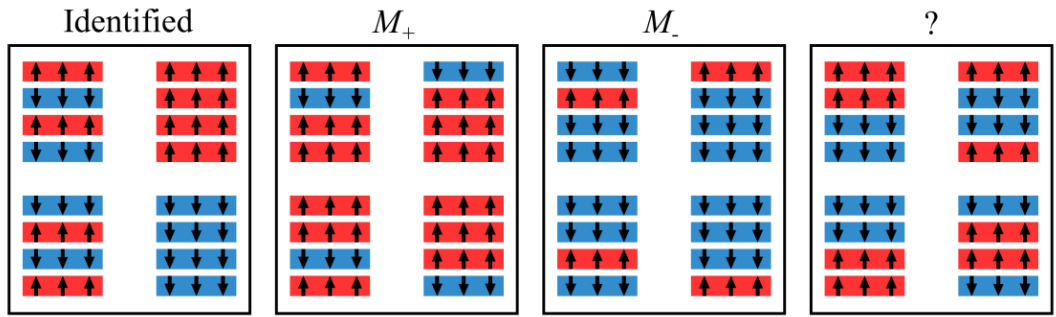

**Figure S9. Possible layered magnetic states for four-layer CrI$_3$.**

We first list all possible 16 magnetic states for four-layer CrI$_3$ in fig. S9, and arrange them into four groups. Among these states, the two layered-antiferromagnetic states (↑↓↑↓ and ↓↑↓↑) and the two fully spin-polarized states (↑↑↑↑ and ↓↓↓↓) have been identified as shown in the first group. Since the intermediate states correspond to the positive and negative RMCD plateaus with half of the height of the two fully spin-polarized states, these states should have three ↑ (↓) and one ↓ (↑) at positive (negative) magnetic fields. Therefore, we identify 8 possible states and group them as $M_+$ {↑↓↑↑,↑↑↓↑,↓↑↑↑ and ↑↑↑↓} at positive fields, and $M_-$ {↓↑↓↓, ↓↓↑↓, ↑↓↓↓ and ↓↓↓↑} at negative fields. If the remaining four states exist in four-layer CrI$_3$, they should correspond to the near-zero RMCD plateau at zero field due to their vanishing net magnetizations. However, these states' energies are higher than the two layered-antiferromagnetic states (↑↓↑↓ and ↓↑↓↑) due to less interfaces between adjacent layers with opposite magnetizations, and thus should not exist at zero field. Furthermore, the very small tunneling current at zero field indicates states with high current-blocking efficiencies. This points to the two layered-antiferromagnetic states (↑↓↑↓ and ↓↑↓↑) as the zero-field ground states, as these two states have more current-blocking interfaces (between adjacent layers with opposite magnetizations) than the remaining four states with zero net magnetization.

**S2.8 Additional four-layer CrI$_3$ sf-MTJs**

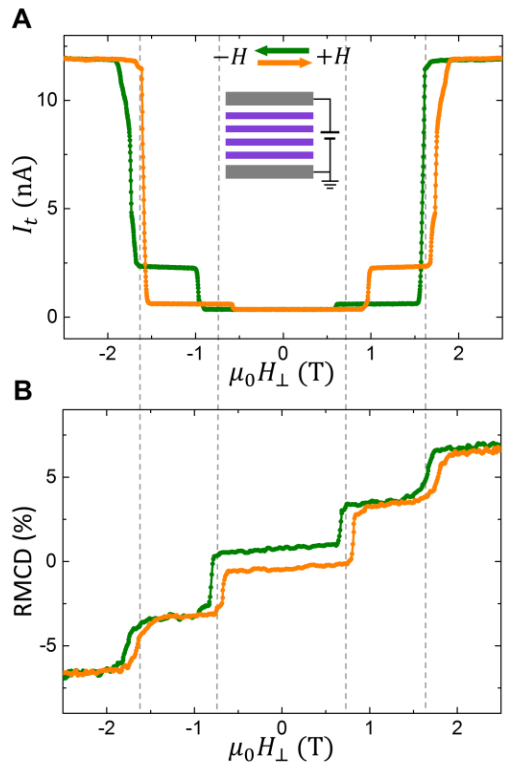

**Figure S10. Tunneling current and RMCD from an additional four-layer CrI$_3$ sf-MTJ.** (**A**) and (**B**) show the tunneling current and RMCD measurements from an additional four-layer CrI$_3$ sf-MTJ.

**References:**

1. P. Blake *et al.*, Making graphene visible. *Appl. Phys. Lett.* **91**, 063124 (2007).

2. B. Huang *et al.*, Layer-dependent ferromagnetism in a van der Waals crystal down to the monolayer limit. *Nature* **546**, 270–273 (2017).

3. P. J. Zomer, M. H. D. Guimarães, J. C. Brant, N. Tombros, B. J. Van Wees, Fast pick up technique for high quality heterostructures of bilayer graphene and hexagonal boron nitride. *Appl. Phys. Lett.* **105**, 013101 (2014).

4. K. Sato, Measurement of magneto-optical kerr effect using piezo-birefringent modulator. *Jpn. J. Appl. Phys.* **20**, 2403–2409 (1981).

5. D. Zhong *et al.*, Van der Waals engineering of ferromagnetic semiconductor heterostructures for spin and valleytronics. *Sci. Adv.* **3**, e1603113 (2017).

6. M. Charilaou, C. Bordel, F. Hellman, Magnetization switching and inverted hysteresis in perpendicular antiferromagnetic superlattices. *Appl. Phys. Lett.* **104**, 212405 (2014).

7. R. R. Birss, *Symmetry and Magnetism* (Elsevier Science & Technology, Amsterdam, North-Holland, 1964).

8. L. Onsager, Reciprocal relations in irreversible processes. I. *Phys. Rev.* **37**, 405–426 (1931).